\documentclass[journal]{IEEEtran}

\usepackage{cite}
\usepackage{graphicx}

\ifCLASSOPTIONcompsoc
  \usepackage[caption=false,font=normalsize,labelfont=sf,textfont=sf]{subfig}
\else
  \usepackage[caption=false,font=footnotesize]{subfig}
\fi

\begin{document}
\title{The Data Acquisition System for the KOTO Experiment
 }
%
%

%
\author{Yasuyuki Sugiyama,
Jia Xu,
Monica Tecchio,
Nikola Whallon,
Duncan McFarland,
Jiasen Ma,
Manabu Togawa,
Yasuhisa Tajima,
Mircea Bogdan,~\IEEEmembership{Member,~IEEE,}
Jon Ameel,
Myron Campbell,
Yau Wai Wah,
Joseph Comfort and 
Taku Yamanaka%
\thanks{Manuscript received June 16, 2014.}

\thanks{This research was supported by KEK, 
MEXT KAKENHI Grant Number 18071006,
JSPS KAKENHI Grant Number 23224007,
the Japan/US Cooperation Program,
the DOE award DE-SC0002644 through a subcontract from the University of Michigan,
and DOE award DE-SC0006497 to Arizona State University.
}%
\thanks{Yasuyuki Sugiyama is with the Department of Physics, Osaka University, Toyonaka, Osaka 560-0043, Japan.
(e-mail: sugiyama@champ.hep.sci.osaka-u.ac.jp)}%
\thanks{Jia Xu is with Department of Physics, University of Michigan, 450 Church St, Ann Arbor, MI 48109, USA
(e-mail: jiaxu@umich.edu)}%
\thanks{Monica Tecchio is with the Department of Physics, University of Michigan, 450 Church St, Ann Arbor, MI 48109, USA.
(e-mail: tecchio@umich.edu)}%
\thanks{Nikola Whallon is with the Department of Physics, University of Michigan, 450 Church St, Ann Arbor, MI 48109, USA.
(e-mail: alokin@umich.edu)}%
\thanks{Duncan McFarland is with the Department of Physics, Arizona State Univeristy, Tempe, AZ 85287, USA.
(e-mail: duncan@azmcfarland.com)}%
\thanks{Jiasen Ma is with the Enrico Fermi Institute, University of Chicago, 5640 South Ellis Ave, Chicago, IL 60637, USA. (e-mail: jsma@uchicago.edu)}%
\thanks{Manabu Togawa is with the Department of Physics, Osaka University, Toyonaka, Osaka 560-0043, Japan.
(e-mail: togawa@champ.hep.sci.osaka-u.ac.jp)}%
\thanks{Yasuhisa Tajima is with the Department of Physics, Yamagata University, Yamagata, Yamagata 990-8560, Japan. (e-mail: tajima@quark.kj.yamagata-u.ac.jp)}%
\thanks{Mircea Bogdan is with the University of Chicago, 5640 South Ellis Ave, Chicago, IL 60637, USA.
(
e-mail: bogdan@edg.uchicago.edu). 
}
\thanks{Jon Ameel is with the Department of Physics, University of Michigan, 450 Church St, Ann Arbor, MI 48109, USA.
(e-mail: sivaluna@umich.edu)}%
\thanks{Myron Campbell is with the Department of Physics, University of Michigan, 450 Church St, Ann Arbor, MI 48109, USA.
(e-mail: myron@umich.edu)}%
\thanks{Yau Wai Wah is with the Enrico Fermi Institute, University of Chicago, 5640 South Ellis Ave, Chicago, IL 60637, USA. (e-mail: ywah@uchicago.edu)}%
\thanks{Joseph Comfort is with the Department of Physics, Arizona State Univeristy, Tempe, AZ 85287, USA.
(e-mail: Joseph.Comfort@asu.edu)}%
\thanks{Taku Yamanaka is with the Department of Physics, Osaka University, Toyonaka, Osaka 560-0043, Japan.
(e-mail: taku@champ.hep.sci.osaka-u.ac.jp)}%
}

\maketitle
\thispagestyle{empty}

\begin{abstract}
We developed and built a new system of readout and trigger electronics,
based on the waveform digitization and pipeline readout,
for the KOTO experiment at J-PARC, Japan.
KOTO aims at observing the rare kaon decay $K_{L}\rightarrow\pi^{0}\nu\bar{\nu}$.
%
%
%

A total of 4000 readout channels from various detector subsystems are digitized by 14-bit 125-MHz ADC modules equipped with a 10-pole Bessel filter in order to reduce the pile-up effects.
The trigger decision is made every 8-ns using the digitized waveform information.
To avoid dead time,
the ADC and trigger modules have pipelines in their FPGA chips to store data while waiting for the trigger decision.

%

The KOTO experiment performed the first physics run in May 2013.
The data acquisition system worked stably during the run.

\end{abstract}

\section{Introduction}
%
%

\subsection{CP Symmetry Breaking}
\IEEEPARstart{W}{hen}
particles and anti-particles were created in the early universe, 
they should have been generated in equal amounts.
However, all the matter around us is made of particles.
This can be explained if the symmetry between the particles and anti-particles,
called  ``CP symmetry",
is broken.
In the Standard Model of particle physics,
the symmetry is broken in the weak interactions for quarks.
The broken symmetry is explained by the Kobayashi-Maskawa model\cite{KM}.
However,
the amount of asymmetry predicted by the model is not large enough to explain the difference in abundance between particles and anti-particles.
This is why we are looking for new physics that can break the CP symmetry.

\subsection{$K_L \to \pi^0 \nu \bar{\nu}$}
$K_L \to \pi^0 \nu \bar{\nu}$ is a rare decay mode of the long-lived neutral K meson, $K_L$.
This decay directly breaks the CP symmetry.

The Standard Model predicts the branching ratio BR$(K_{L} \to \pi^0 \nu \bar{\nu})$ to be $(2.43\pm0.06)\times10^{-11}$\cite{bros},
while the current experimental upper limit is $2.6\times10^{-8}$ at the 90\% confidence level (C.L.) from the KEK E391a experiment\cite{e391afinal}.
The theoretical uncertainty on the prediction is around 2\%.
A a measurement of the branching ratio different from the theoretical prediction
would signify new physics beyond the Standard Model.

A model-independent upper bound on  BR($K_L \to \pi^0 \nu \bar{\nu}$),
called the Grossman-Nir bound\cite{GN97}, is derived as
\[
BR(K_{L} \to \pi^0 \nu \bar{\nu}) < 4.4 \times BR(K^{+} \to \pi^+ \nu \bar{\nu}). 
\]
This relation between the decay modes of neutral and charged kaons is obtained from the isospin symmetry arguments.
The measured BR($K^{+} \to \pi^+ \nu \bar{\nu}$) at (1.7 $\pm$ 1.1) $\times 10^{-10}$ from the BNL E787 and E949 experiments\cite{bnl949}
yields an upper limit on BR($K_L \to \pi^0 \nu \bar{\nu}$) of 1.2 $\times10^{-9}$ (68 \% C.L.).

Some theoretical models of new physics enhance BR$(K_{L} \to \pi^0 \nu \bar{\nu})$.
They introduces SUSY particles\cite{BBIL05},
littlest Higgs\cite{blankeLHT,gotoLHT} and fourth generation quarks and leptons\cite{hou,buras2010}.
Improving the upper limit on the branching ratio can put constraints on such new physics models.


\subsection{KOTO experiment}
The  KOTO experiment\cite{proposal} is a new kaon-decay experiment at J-PARC\cite{jparc} in Ibaraki, Japan.
Its goal is to discover the $K_L \to \pi^0 \nu \bar{\nu}$ decay
and measure its branching ratio.
KOTO is designed to improve the sensitivity of E391a by three orders of magnitude.
The experimental difficulty in observing this decay is due to the lack of charged particles in the final state 
and to the high efficiency required for photon detection.

The strategy of KOTO 
is to detect ``$\pi^{0} $ plus nothing'',
where "Nothing" refers to the difficulty to detect $\nu$'s (neutrino).
Other decay modes of $K_L$ such as  $K_L \to 3\pi^0 $,
$K_L \to 2\pi^0 $,
and $K_L \to \pi^+\pi^-\pi^0 $ can be sources of background 
when the extra $\pi^0$'s or charged particles in the final states are not observerd.

\begin{figure}[htb]
	\begin{center}
\includegraphics[width=\linewidth,clip,bb=0 0 740 320]{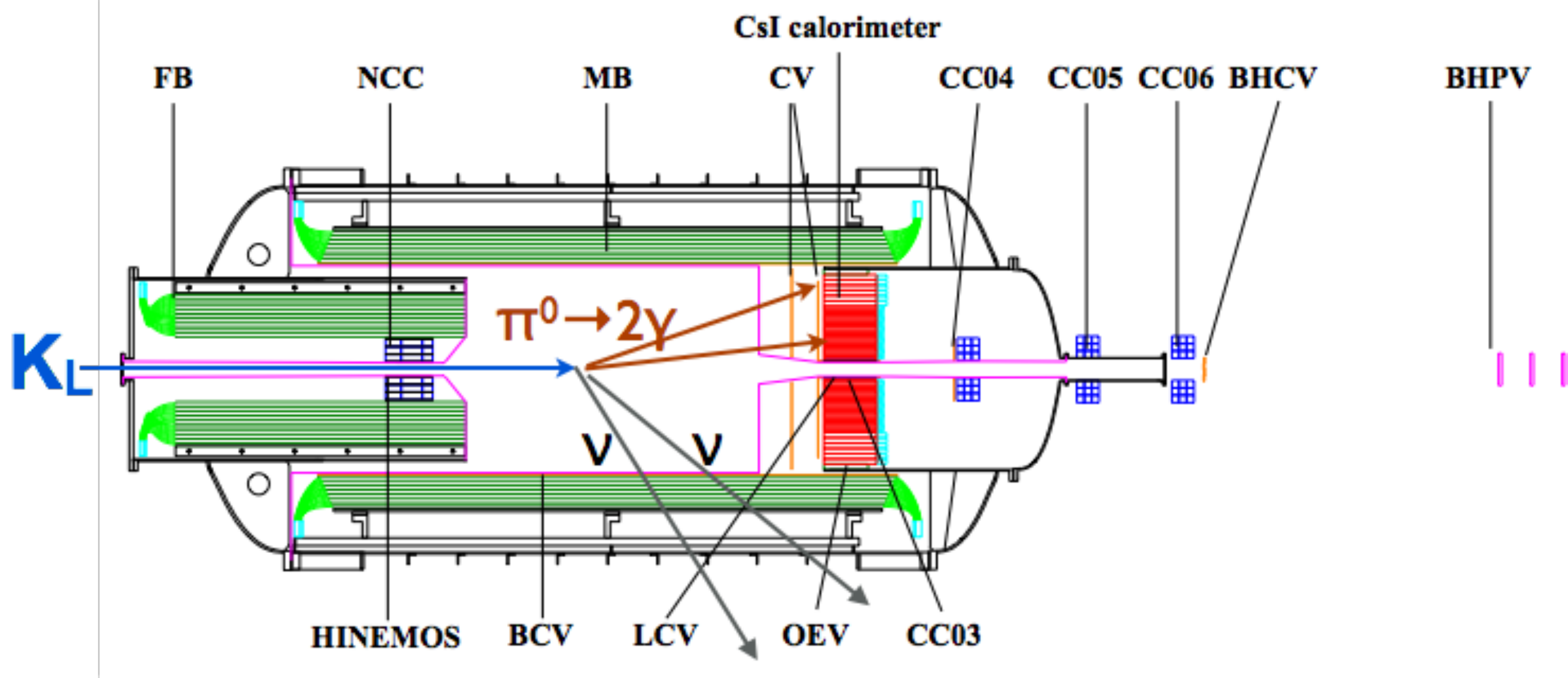}
	\end{center}
	\caption{Sideview of the KOTO detector.}
		\label{fig:detector}
	
\end{figure}

Figure \ref{fig:detector} shows the detecter of the KOTO experiment.
A neutral beam enters into the KOTO detector.
The two photons from $\pi^{0}$ in $K_L \to \pi^0 \nu \bar{\nu}$ are detected by an electromagnetic calorimeter made of undoped Cesium Iodide(CsI) crystals.
We also require no activity in other detector subsystems surrounding the decay region to discard background events due to other $K_{L}$ decay modes.
We will refer to these subsystems as "Veto Detectors".

The KOTO experiment performed its first physics run for five days in May 2013.

\section{The Data acquisition system for the KOTO experiment}
The requirements for the data acquisition system are 14-bit dynamic range for the energy measurement and sub-nanosecond timing resolution to resolve overlapping events. 
A data acquisition without dead time is also required to cope with the high counting rates expected for the detector in the high-intensity beam from J-PARC.

To satisfy these requirements,
we designed a new system of readout and trigger electronics based on waveform digitization and pipeline readout\cite{daq_monica}.
%

A total of 4000 channels of output signals from the KOTO detector subsystems are digitized with 14-bit 125-MHz ADC modules and are stored in buffers inside Field Programmable Gate Arrays (FPGAs) until, a trigger decision is  made. 
The trigger system,
with three levels,
uses the waveform information
to make a trigger decision of increased sophistication at each level.

\begin{figure}[htb]
	\begin{center}
\includegraphics[width=\linewidth,clip,bb=0 0 1038 460]{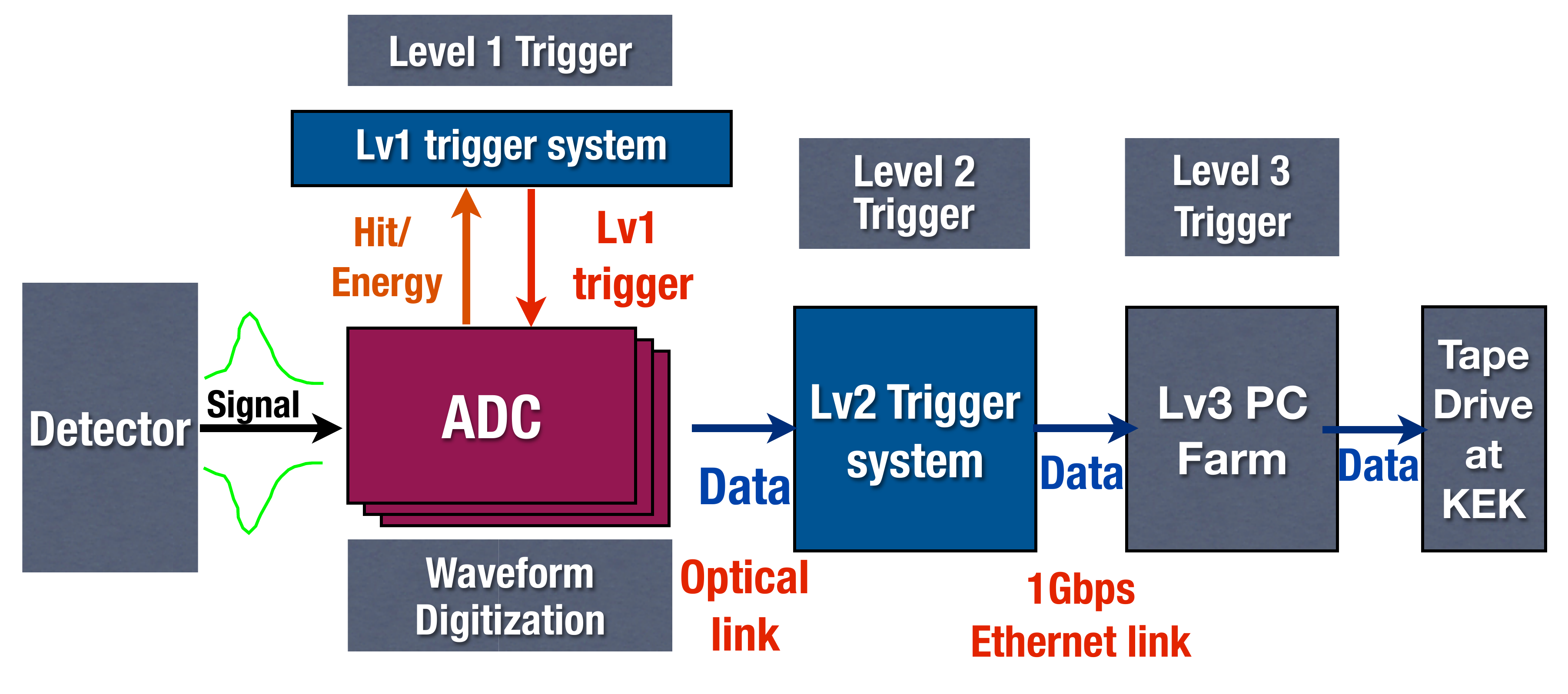}

	\end{center}
	\caption{Schematic view of the KOTO data acquisition system.}
		\label{fig:daq_overview}
	
\end{figure}

Figure \ref{fig:daq_overview} shows a schematic view of the KOTO data acquisition system.
\subsection{Waveform Digitization}
\begin{figure}[htb]	
	\begin{center}
\includegraphics[width=\linewidth,clip,bb= 0 0 1085 586]{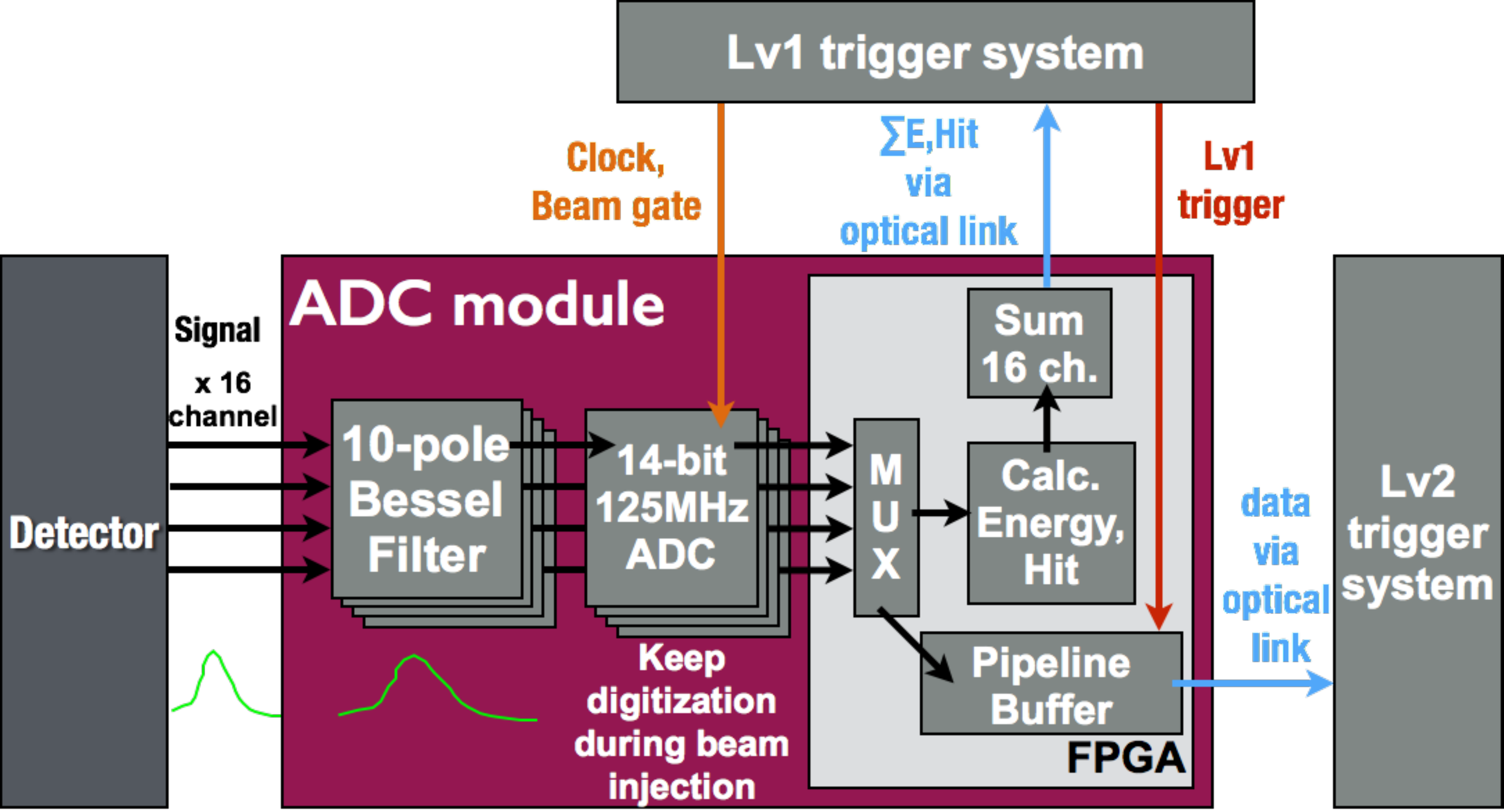}
	\end{center}
	\caption{ADC module with a 10-pole Bessel Filter.}
\label{fig:fadcscheme}
\end{figure}

\begin{figure}[htb]	
	\begin{center}
\includegraphics[width=\linewidth,clip,bb=50 0 1020 381]{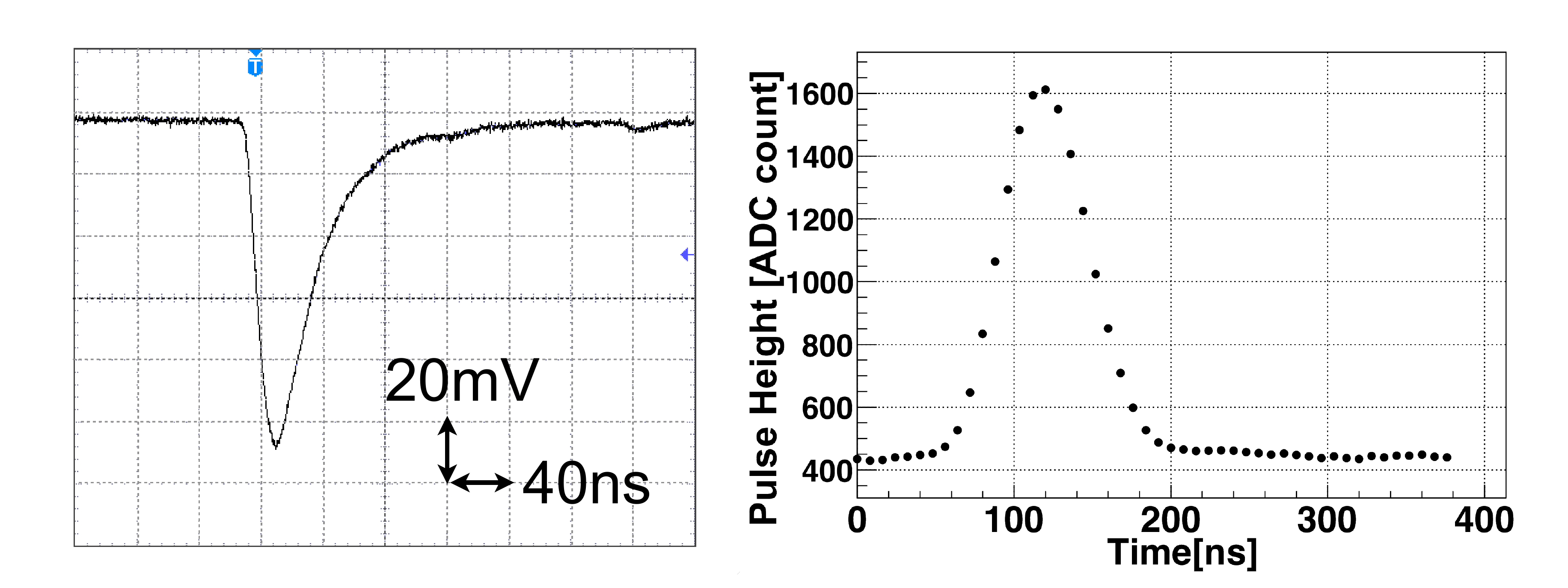}
	\end{center}
	\caption{Signal from a CsI crystal with a photomultiplier tube recorded by an oscilloscope (left);
	recorded pulse shape by a 125MHz ADC with a 10-pole Bessel filter (right).}
\label{fig:waveform}
\end{figure}

We designed new 125-MHz ADC modules \cite{mircea1,mircea2} for the fronted electronics of KOTO to record the waveform from each detector channel.
Figure \ref{fig:fadcscheme} shows a schematic view of the ADC module.
The inputs to the ADC chip are widened to be in a gaussian-shape with a 10-pole Bessel filter.
Figure \ref{fig:waveform} shows the signal from a CsI crystal  with a photomultiplier tube, as recorded by an oscilloscope, and the relative pulse shape as recorded by a 125-MHz ADC module with the filter.
The leading edge of photomultiplier signal is too fast for an 8-ns sampling.
Shaping the signal into a gaussian-shape pulse can effectively increase the number of sampling points in the leading edge. 
With this technique,
a timing resolution around 1-ns can be achieved even with an 8-ns sampling.
To avoid dead time,
the digitized waveform data are stored in 4$\mu$s long pipelines inside the ADC modules while waiting for the trigger decision. 
The data output from each ADC module is sent to the trigger system via 2.5-Gbps optical links. 


\subsection{Trigger System}

The KOTO trigger system consists of three levels.
The first two levels are implemented in custom designed hardware modules. 
The third level consists of a computer farm.

\subsubsection{Lv1 trigger system}
\begin{figure}[htb]	
	\begin{center}
\includegraphics[width=\linewidth,clip,bb=0 0 683 412]{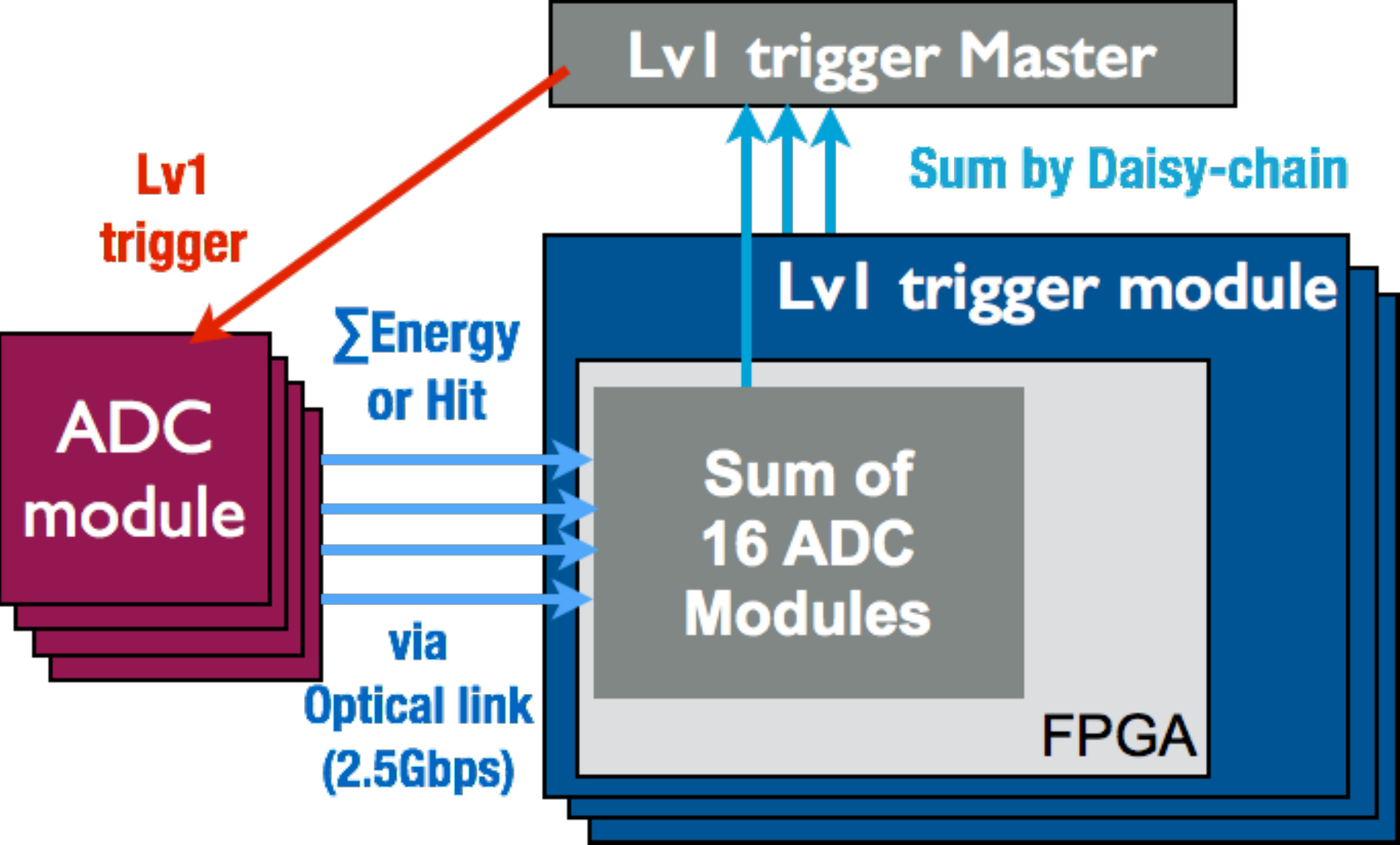}
	\end{center}
	\caption{Lv1 trigger system with the optical link and daisy-chain.}
\label{fig:lv1scheme}
\end{figure}

The first-level (Lv1) trigger decision is made every 8-ns,
by requiring a minimum energy in the CsI calorimeter and no activities in the Veto detectors.
Figure \ref{fig:lv1scheme} shows a schematic view of the Lv1 trigger system.
One of the Lv1 trigger modules receives information from sixteen ADC modules via optical fibers.
The data is summed up inside each Lv1 trigger module first
and subsequently summed over all the Lv1 trigger modules via a daisy-chain bus.
The Lv1 trigger master module receives the total energy in the CsI calorimeter
together with a summary of the activity in all the other detector subsystems,
and makes the Lv1 trigger decision.
The decision is broadcasted to each ADC module which,
in turn,
saves the event data as it exits the pipeline and send it to the second-level trigger module for further trigger decision.

\subsubsection{Lv2 trigger system}
The second-level (Lv2) trigger decision is made based on the recorded waveform information for the event.
Calculation of the "Center of Energy (COE)" and of the number of photons in the CsI calorimeter can be implemented at this stage.
The COE is used to select the events with a large transverse momentum,
$p_{T}$.
Due to the neutrinos in the final state,
the $K_{L}\rightarrow \pi^{0} \nu \bar{\nu}$ decay event should have a large $p_{T}$ compared to most of the background sources.  
The number of photon can be used to distinguish $K_{L}\rightarrow \pi^{0} \nu \bar{\nu}$ from other decay modes with multiple $\pi^{0}$'s such as $K_L \to 3\pi^0 $ and $K_L \to 2\pi^0 $. 
At present,
only the COE calculation is implemented in the Lv2 trigger system.

\begin{figure}[htb]	
	\begin{center}
\includegraphics[width=\linewidth,clip,bb=0 0 965 469]{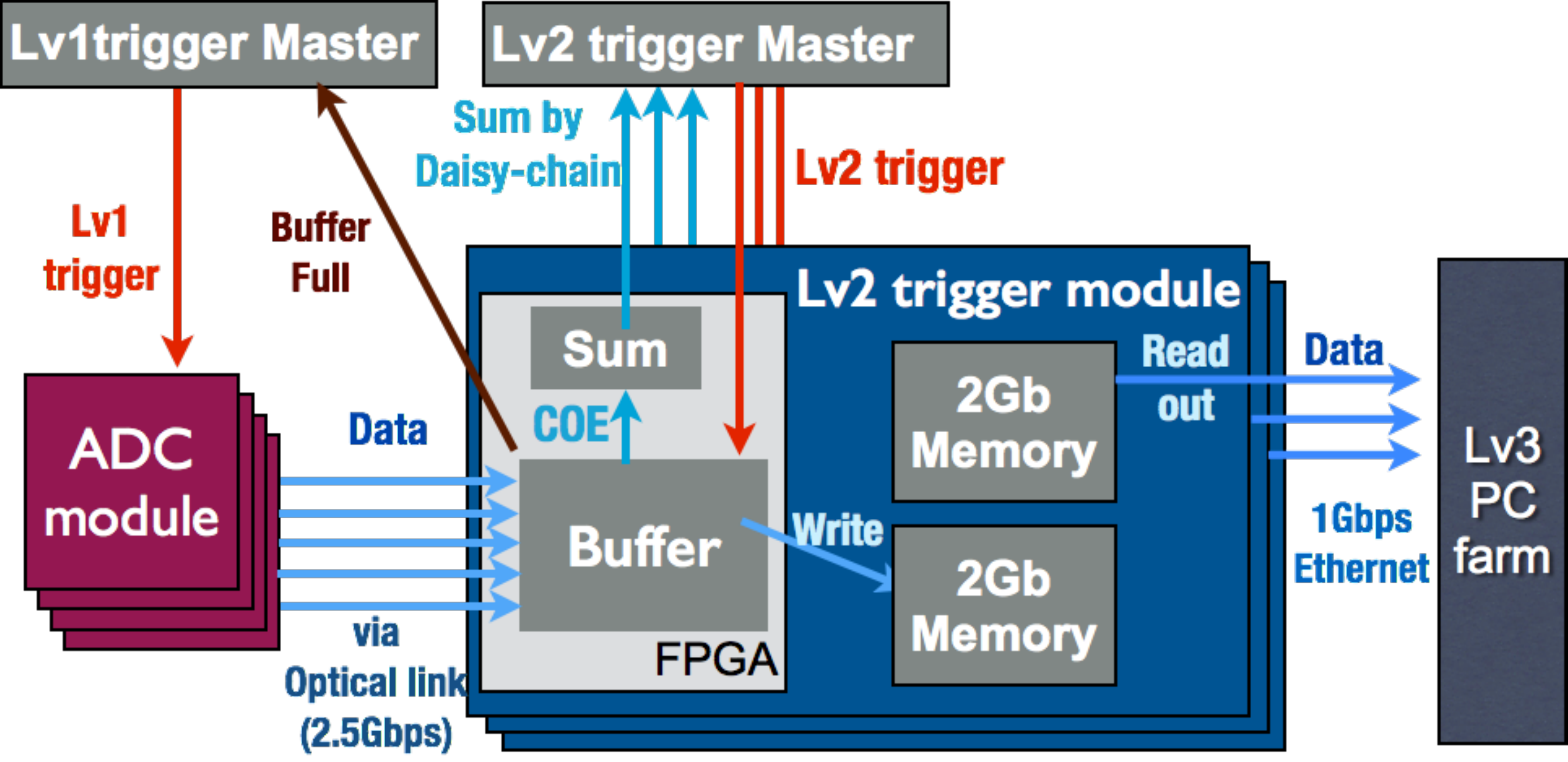}
	\end{center}
	\caption{Lv2 trigger system, which receives data from ADC modules via the optical links and sends the data to the Lv3 PC farm via the 1-Gbps ethernet.}
\label{fig:lv2scheme}
\end{figure}
Figure \ref{fig:lv2scheme} shows a schematic view of the Lv2 trigger system.
The Lv2 trigger system receives data packets of a digitized waveforms from sixteen ADC modules via the optical fibers.
The data packets are stored on a buffer while waiting for the Lv2 trigger decision.
When this buffer becomes full, the Lv1 trigger decision is suspended.
This in turn causes dead time for the whole trigger system.

The COE for the full calorimeter is calculated via the daisy-chain bus in the same fashion as for Lv1 trigger system.
The Lv2 trigger master compares the total COE to a minimum value and generate the Lv2 trigger decision. 
Data for events passing the Lv2 decision is stored on 2-Gbit memory and are eventually transferred
to a computer farm via a 1-Gbps Ethernet.
The Lv2 trigger module uses two 2-Gbit memories to write events into one while reading from the other. 

\subsubsection{Lv3 Trigger system}
The third-level (Lv3) trigger decision is made by a computer farm.
The computer farm collects the data packets from the Lv2 trigger modules and builds the events.

\begin{figure}[htb]	
	\begin{center}
\includegraphics[width=\linewidth,clip,bb=0 0 855 428]{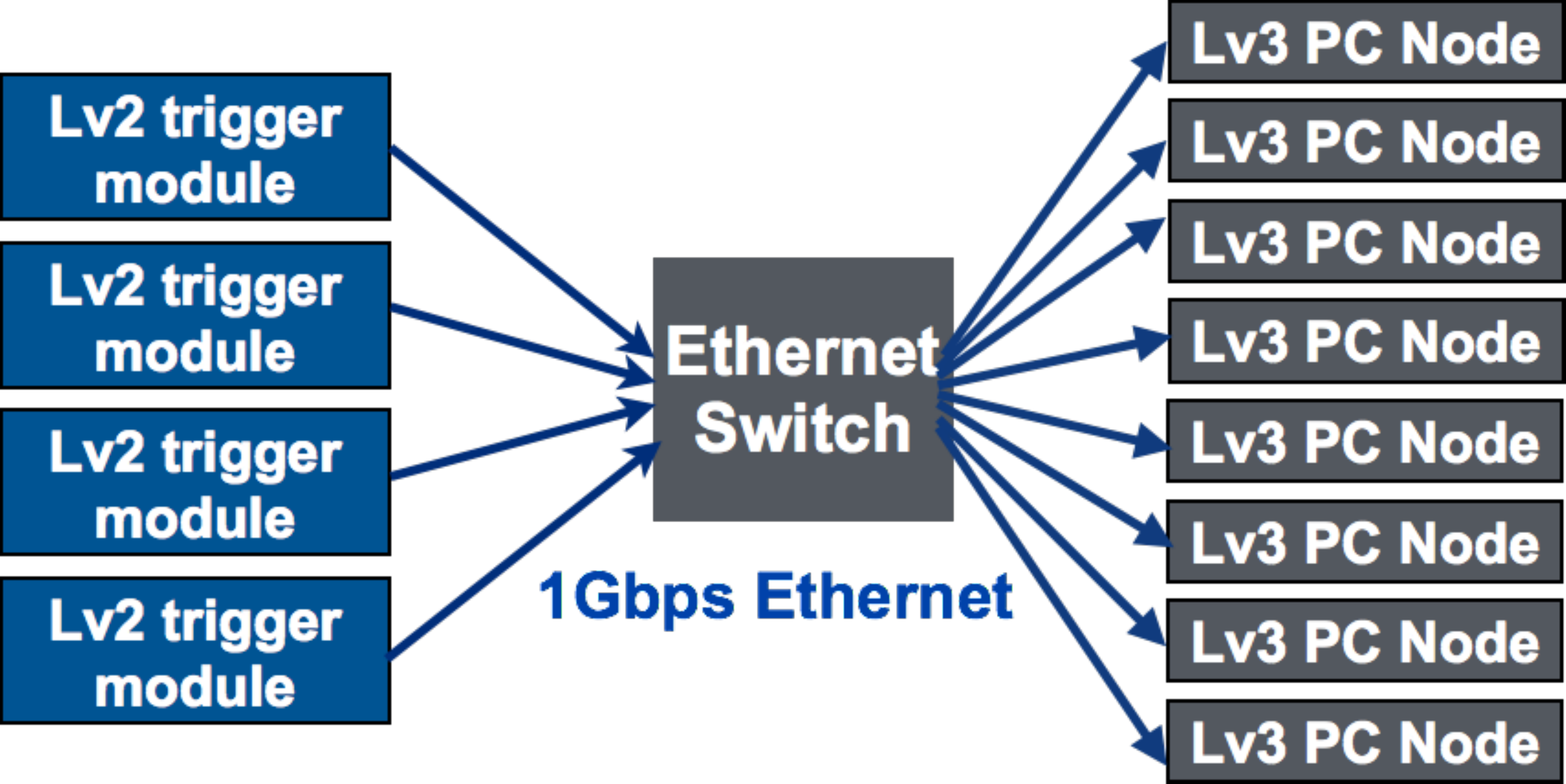}
	\end{center}
	\caption{Event building by the Ethernet switch.}
\label{fig:lv3scheme}
\end{figure}
Figure \ref{fig:lv3scheme} shows how events are built at the Lv3 trigger system.
The Lv2 trigger modules change the destination of the data packets for each event,
 and send them to an Ethernet switch.
The Ethernet switch transfers data packets for different events to different Lv3 PC nodes.
Thus each node receives the full information for a given events.

The Lv3 trigger decision is made using fully reconstructed events. 
Data selection with a more detailed waveform analysis using fitting,
as well as data compression,
can be implemented at this stage.
At present, only the data compression is implemented.
The events which passed all the trigger decisions are 
transferred from the J-PARC site in the Tokai village
to the storage in the KEK Computer Research Center in Tsukuba.

\section{DAQ Performance in the 2013 Physics Run }
In this section,
the operation and performance of the KOTO data acquisition system during the physics run in May 2013 are described.

\subsection{DAQ Stability}
\begin{figure}[htb]	
	\begin{center}
\includegraphics[width=\linewidth,clip,bb=0 0 780 381]{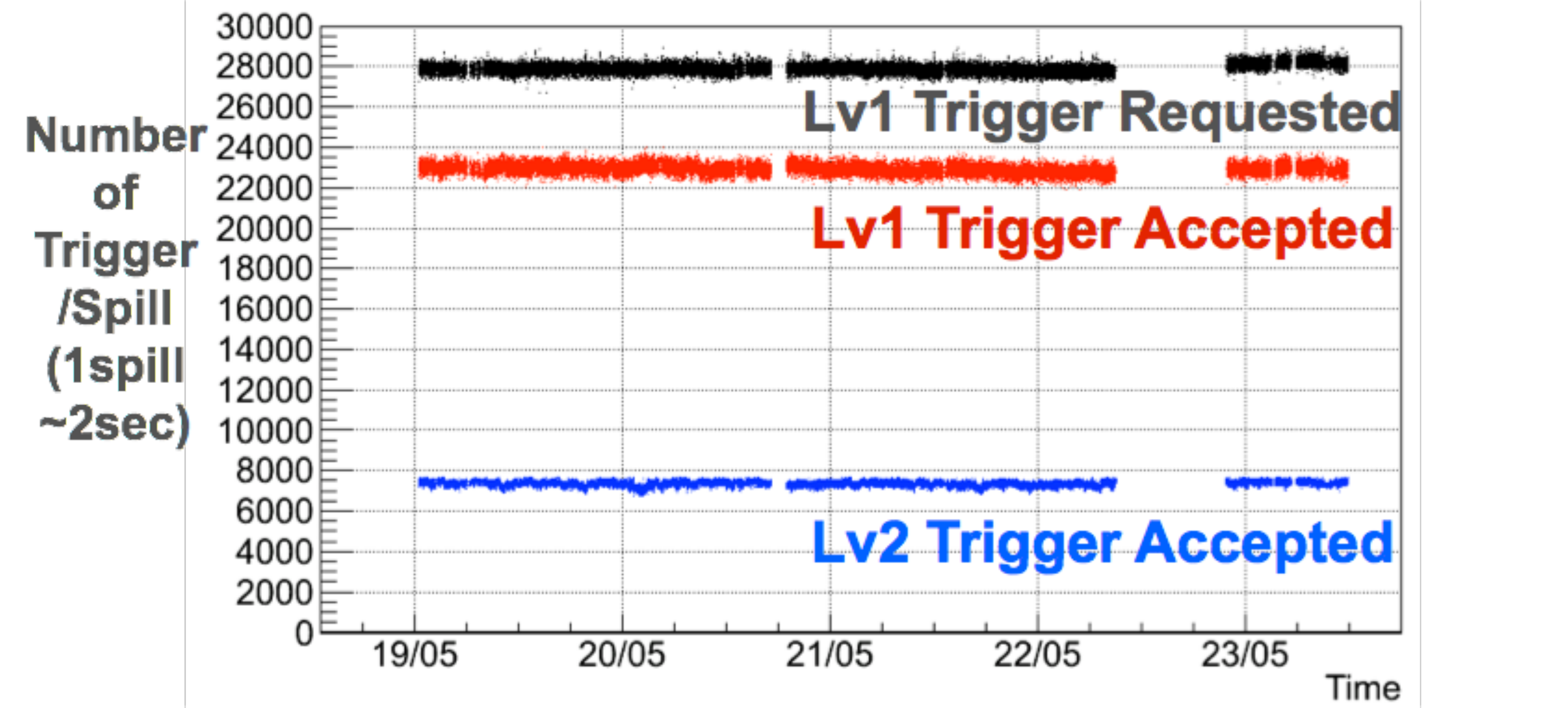}
	\end{center}
	\caption{Number of triggers in each spill during the physics run in May 2013.}
\label{fig:stability}
\end{figure}

Figure \ref{fig:stability} shows the numbers of triggers in each beam spill\footnote{A beam extraction from the accelerator is called ``spill". The beam is extracted for the 2-sec  every 6-sec.}.
The number of triggers requested and accepted at the Lv1 and Lv2 trigger stages were stable during the physics run. 
The number of Lv1 triggers accepted was 17 \% smaller than the number of Lv1 triggers requested; 
this is due to the dead time incurred by trigger system as a result of the limited  buffering available inside the Lv2 trigger module.

\begin{figure}[htb]	
	\begin{center}
\includegraphics[width=\linewidth,clip,bb=0 0 792 612]{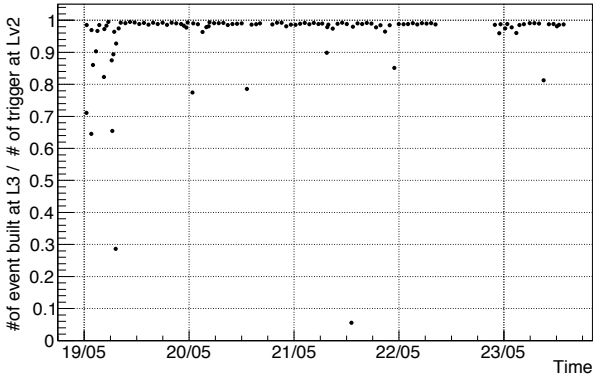}
	\end{center}
	\caption{Fraction of the number of the events accepted at the Lv2 and build at Lv3 for each data-taking run.}
\label{fig:l3ratio}
\end{figure}

Figure \ref{fig:l3ratio} shows the fraction of the number of events accepted at the Lv2 and build at Lv3 stages for each data-taking run.
Event loss during the data transmission from the Lv2 to the Lv3 trigger system is only a few percent for most of the runs.
The average fraction of events built at Lv3 was 97\%.

\subsection{Trigger Performance}
\subsubsection{Lv1 trigger}
The Lv1 trigger decision is made from the total energy recorded in the CsI Calorimeter and in the other Veto detectors.
The information from the CsI calorimeter is used to select events with energy deposition consistent with the presence of a $\pi^{0}$ from a $K_{L}$ decay while the information from the other detector subsystems
is used to remove background events.

\begin{figure}[htb]	
	\begin{center}
\includegraphics[width=.9\linewidth,clip,bb= 0 0 1024 768]{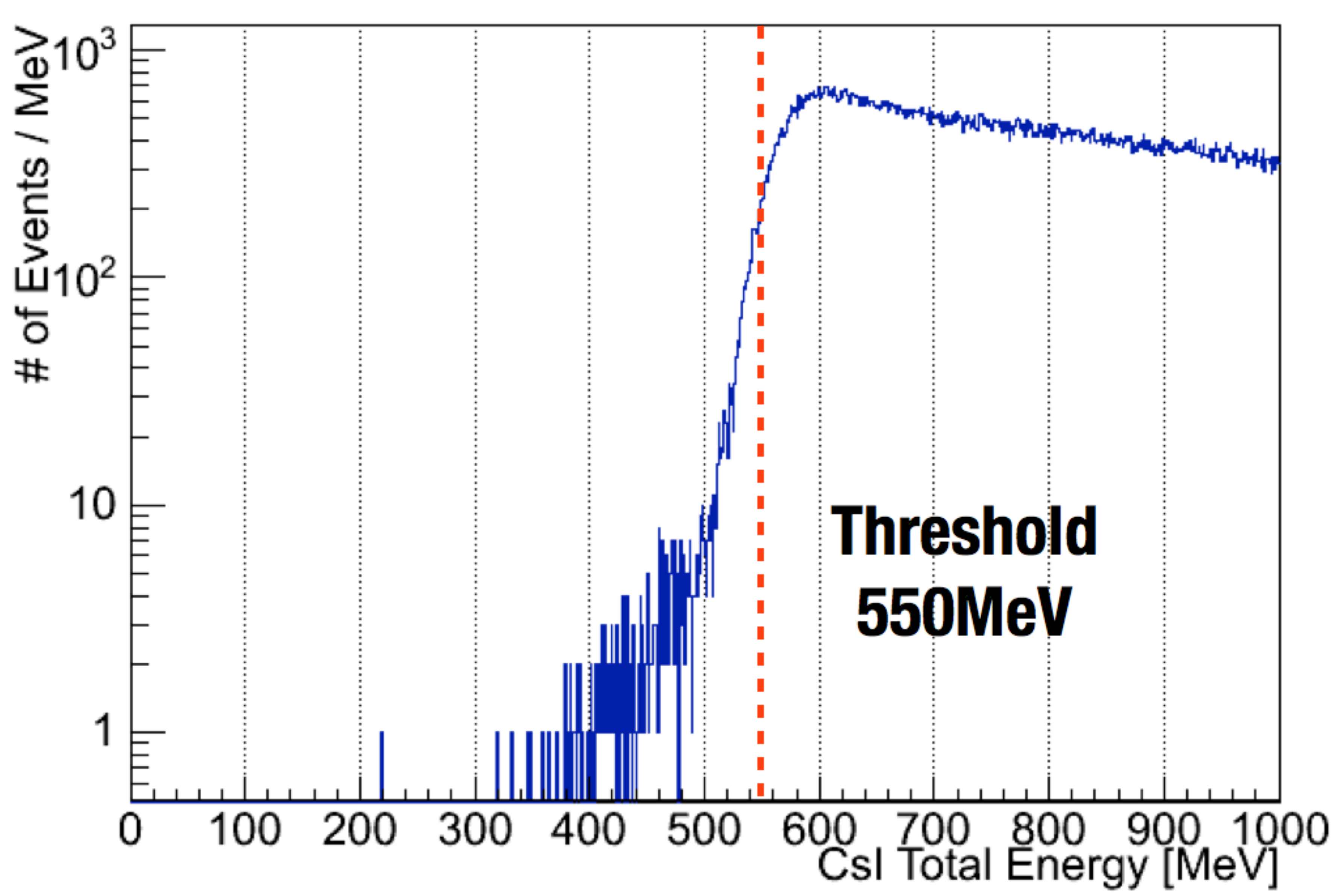}
	\end{center}
	\caption{Total energy in the CsI calorimeter;
	 the dotted line represents the trigger threshold.}
\label{fig:lv1csi}
\end{figure}

Figure \ref{fig:lv1csi} shows the total energy in the CsI calorimeter for the events passing  the Lv1 trigger decision.
The trigger threshold was set to be at 550 MeV;
the number of triggers was 258k triggers/spill.

\begin{figure}[htb]	
	\begin{center}
\includegraphics[width=\linewidth,clip,bb=  0 0 1024 701]{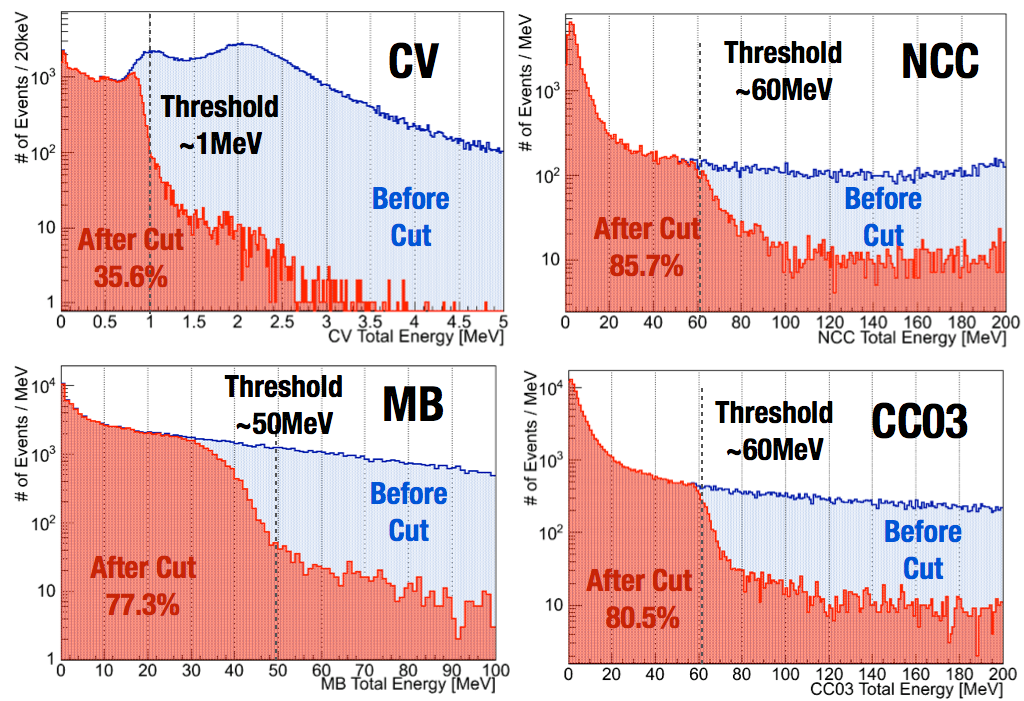}
	\end{center}
	\caption{Visible Energy in the veto subsystems;
	 the dotted lines represent the trigger thresholds.
	 The number of "After Cut" in each histogram shows the ratio of the hits of each subsystem before and after the Lv1 trigger is imposed.}
\label{fig:lv1veto}
\end{figure}

Figure \ref{fig:lv1veto} shows the energy in the veto subsystems before and after the trigger is imposed.
A clear edge is seen around the energy threshold required on the Lv1 trigger decision.
After imposing these Veto cuts, 
the number of triggers was reduced to 27k triggers/spill.
%
%
\subsubsection{Lv2 trigger}
\begin{figure}[htb]	
	\begin{center}
\includegraphics[width=\linewidth,clip,bb= 0 0 1024 753]{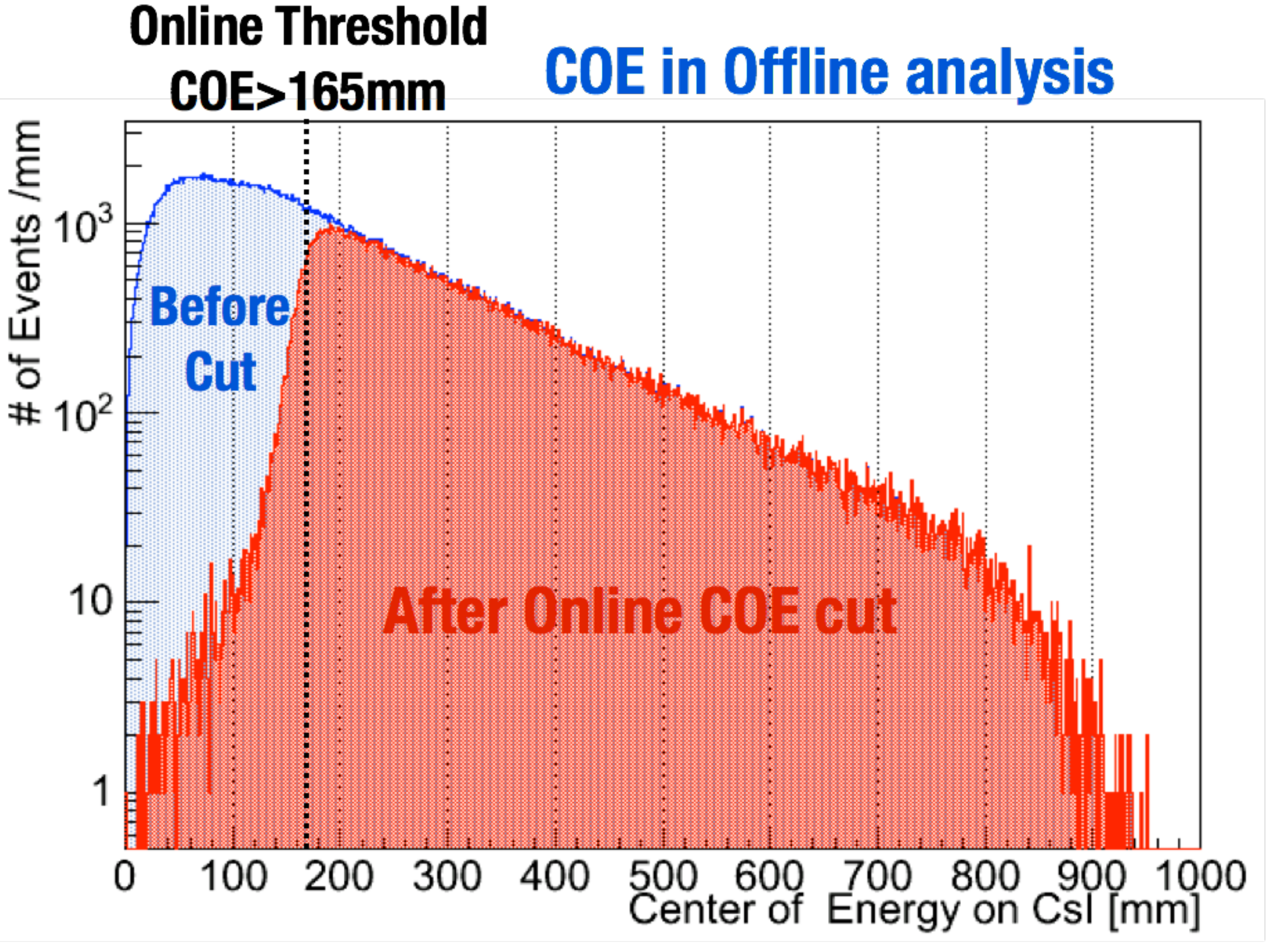}
	\end{center}
	\caption{Center of Energy (COE) in the CsI Calorimeter;
	the dotted line represents the trigger threshold.}
\label{fig:lv2coe}
\end{figure}

At the Lv2 trigger stage, the COE in the CsI calorimeter is calculated and used to make the trigger decision.
To select events with a large $p_{T}$,
we required the COE in the CsI calorimeter to be larger than 165mm. 
Figure \ref{fig:lv2coe} shows the COE distribution with and without this requirement. 
About 6.8\% of the events with high COE after event reconstruction are lost due to the online trigger COE cut but they all have a reconstructed COE lower than the threshold used in the offline analysis.
After imposing COE cuts,
the number of triggers was reduced to 8k triggers/spill.

\subsubsection{Lv3 trigger}
The Lv3 trigger decision is made in software inside the Lv3 PC farm.
No event selection was made at the Lv3 stage in the physics run in May 2013;
only the data compression was applied to reduce the data size by a factor of 3.
The total data-transfer speed from the Lv2 trigger system to the Lv3 PC farm was 5-Gbps,
while the maximum data-transfer speed from the Lv3 PC farm to the KEK Tsukuba campus was 3-Gbps.
With this data compression,
we were able to transfer data from Tokai to the  tape storage system in Tsukuba without any event loss.

\section{discussion}
For the next KOTO physics run,
the KOTO DAQ system will be upgraded as to reduce the two main sources of event loss.
The event loss at Lv1 trigger level will be mitigated by applying 
data compression or 
zero-suppression inside the ADC module,
which greatly reduces the dead time at the Lv2 trigger stage.
An event building with faster PC nodes and a faster bus between nodes will reduce the packet loss during the event building,
and will provide extra processing power for further event selection.
Implementation of event selection at Lv3 will be necessary to reduce the amount of data to be stored on tape.

\section{Conclusion}
We developed and built a new system of readout and trigger electronics based on waveform digitization and pipeline readout for the $K_{L}\rightarrow \pi^{0} \nu \bar{\nu}$  experiment at J-PARC.
It worked stably during the first physics run in May 2013.

\section*{Acknowledgment}
We would like to express our gratitude to the KEK Computer Research Center and the Network and Computing Service Center of Yamagata University for the computer and network resources.
We would like to thank the National Institute of Informatics (NII) for the SINET4 network support. 
We also would like to appreciate all members of the J-PARC accelerator group and hadron beam group for their supports.
%


%
%
%
%
%

\end{document}